\pdfoutput=1
\documentclass[twocolumn,times]{aastex63}

\usepackage{amsmath}

\shorttitle{The Causes of Peripheral Coronal Loop Contraction and
Disappearance}
\shortauthors{Wang et al.}

\begin{document}

\title{The Causes of Peripheral Coronal Loop Contraction and
  Disappearance Revealed in a Magnetohydrodynamic Simulation of Solar
  Eruption}

\correspondingauthor{Chaowei Jiang}
\email{chaowei@hit.edu.cn}

\author[0000-0001-9268-2966]{Juntao Wang}
\affiliation{Institute of Space Science and Applied Technology, Harbin Institute of Technology, Shenzhen 518055, People's Republic of China}
\affiliation{Nanchang Institute of Technology, Nanchang 330044, People's Republic of China}

\author[0000-0002-7018-6862]{Chaowei Jiang}
\affiliation{Institute of Space Science and Applied Technology, Harbin Institute of Technology, Shenzhen 518055, People's Republic of China}

\author[0000-0002-9514-6402]{Ding Yuan}
\affiliation{Institute of Space Science and Applied Technology, Harbin Institute of Technology, Shenzhen 518055, People's Republic of China}

\author[0000-0002-8474-0553]{Peng Zou}
\affiliation{Institute of Space Science and Applied Technology, Harbin Institute of Technology, Shenzhen 518055, People's Republic of China}

\begin{abstract}
  The phenomenon of peripheral coronal loop contraction during solar
  flares and eruptions, recently discovered in observations, gradually
  intrigues solar physicists. However, its underlying physical
  mechanism is still uncertain. One is \cite{hud2000}'s implosion conjecture
  which
  attributes it to magnetic pressure reduction in the magnetic energy
  liberation core, while other researchers proposed alternative
  explanations.
  In previous observational studies we also note the disappearance of
  peripheral shrinking loops in the late phase, of which there is a
  lack of investigation and interpretation. In this paper, we exploit a full
  MHD simulation of solar eruption to study the causes of the two
  phenomena. It is found that the loop motion in the periphery is well
  correlated with magnetic energy accumulation and dissipation in the
  core, and the loop shrinkage is caused by a more significant
  reduction in magnetic pressure gradient force than in magnetic
  tension force, consistent with the implosion conjecture. The
  peripheral contracting loops in the late phase act as inflow to
  reconnect with central erupting structures, which destroys their
  identities and naturally explains their disappearance. We also
  propose a positive feedback between the peripheral magnetic
  reconnection and the central eruption.
\end{abstract}


\section{Introduction} \label{sec:intro}

Solar flares and eruptions are the most violent activities releasing
enormous energy in the solar system. The liberated energy, in the form
of radiation, plasma thermal and nonthermal energy, and bulk kinetic
energy, is believed to originate from free magnetic energy stored in
coronal nonpotential magnetic field. Because of the equivalence
between magnetic energy density and pressure, solar events could cause
a significant reduction in magnetic pressure in the energy releasing
core. Based on this argument, \cite{hud2000} conjectured a phenomenon
termed coronal implosion, which predicts that the plasma and entrained
magnetic field would collapse into the region of reduced magnetic
pressure, and coronal loop contraction may be one of manifestations.

\cite{aly1984,aly1991} and \cite{stu1991} demonstrated that for simply
connected force-free field with a given magnetic flux distribution at
the bottom boundary, the open-field configuration in which all the
field lines are opened to infinity possesses the maximum energy. It
implies that closed field could not completely open through ideal MHD
processes spontaneously without extra energy input. However, in
observations filament eruptions without flares are prevalent
\cite[e.g.,][]{zou2019b}. To resolve this contradiction,
\cite{stu2001} proposed a metastable flux rope model in 3D where the
flux rope is anchored below a restraining arcade and can erupt with
the overlying field partially open. Following MHD simulations which
exploited this magnetic configuration of \cite{stu2001} showed that
during the eruption, part of the unopened arcade field straddling the
two legs of the flux rope would finally shrink toward the erupting
core \citep{aul2005,gib2006b,fan2007,rac2009}.

Though the theoretical argument and simulations both support the
existence of the phenomenon of coronal loop contraction in solar
energy releasing events, its observational evidence is rare so far,
compared to massive samples of solar eruptions and
flares. Contractions of coronal loops with a face-on geometry were
observed at the periphery of active regions in extreme ultraviolet in
a few events
\citep{liu2009,gos2012,liu2012,sun2012,sim2013,yan2013,kus2015,wang2016}. It
was in doubt that these apparent loop contractions may just be a
projection effect as accompanied eruptions could push peripheral loops
to incline. Then \cite{pet2016} and \cite{wang2018} reported edge-on
coronal loop contractions, directly substantiating the reality of the
contracting motion. Especially, \cite{wang2018} showed in two events
that without significant eruptions, loops could still shrink toward
flaring regions. \cite{sim2013} found a good correlation between loop
contraction speed and radiation in hard X-ray and microwave.

In some events, considerable oscillations of the contracting loops (or
part of the contracting loops) were also observed
\citep{liu2010,gos2012,liu2012,sun2012,sim2013}. \cite{liu2010}
suggested that magnetic strands with different contracting speeds may
interact to produce oscillations. \cite{rus2015} proposed a model
where one loop is regarded as a harmonic oscillator with its
equilibrium position changing due to underlying magnetic energy
release, trying to provide an unified interpretation of loop
contractions and oscillations. \cite{sar2017} implemented the first
MHD simulation concentrating on implosion, which found that loops in
regions of different plasma $\beta$ may exhibit different dynamic
properties, and both kink and sausage modes of oscillations could
exist when loops contract. \cite{pas2017} took loop shrinkage into
account in coronal seismology analysis, and only found the fundamental
kink mode for the event in \cite{sim2013}.

Although the implosion phenomenon was conjectured by \cite{hud2000}
as due to magnetic pressure reduction in magnetic energy liberation
region, \cite{zuc2017} and \cite{dud2017} proposed alternative
physical explanations for the peripheral coronal loop contraction. They
attributed the contracting motion of the loops around the periphery to the
return part of a
vortex flow. However, in order to study the change in loop motion, either from
a static state to contraction or from expansion to contraction, the acting
forces should be analysed. Thus in terms of forces, \cite{zuc2017} eliminated
magnetic pressure or energy reduction below the contracting loops as a
candidate, which results in the only possibility of enhanced magnetic tension
in a zero-$\beta$ MHD simulation. Especially, in Sections 3.2.2 and 3.2.3 of
\cite{zuc2017}, ``generates magnetic tension" or ``magnetic tension increases" was
frequently invoked. Though ``overpressure" was also used in Section 3.2.3 of \cite{zuc2017},
their Figure 9 shows that the magnetic pressure gradient is outward rather than inward, which thus could not
explain the loop contraction. They also exploited an analogy between \textbf{the}
hydrodynamic and MHD cases for vortex generation, but the
underlying forces in these two situations are very different. In hydrodynamic
case, the fluid pressure is predominant, and the vortex generation is related
to viscosity, while in coronal MHD the Lorentz force or its two components,
i.e.,
magnetic tension force and magnetic pressure gradient force, dominate, and
viscosity can be neglected in the simulation.
Therefore, the underlying physics for peripheral coronal
loop contraction is still controversial. In addition, we note a
phenomenon of disappearance of shrinking loops (or part of shrinking
loops) in previous studies
\citep{liu2009,sun2012,sim2013,yan2013,kus2015,pet2016,wang2016,wang2018},
which has not been investigated and interpreted so far.

In this paper, we will utilize an MHD simulation to both qualitatively
and quantitatively study the dynamics of peripheral contracting loops,
trying to reveal the physical mechanisms of their motion and
disappearance. The MHD model is described in Section~\ref{sec:method},
and the analysis of peripheral contracting loops is detailed in
Section~\ref{sec:results}. Discussion and conclusions are presented in
Section~\ref{sec:conclusions}.

\begin{figure*}
  \begin{centering}
    \hspace{0cm}\includegraphics[scale=1]{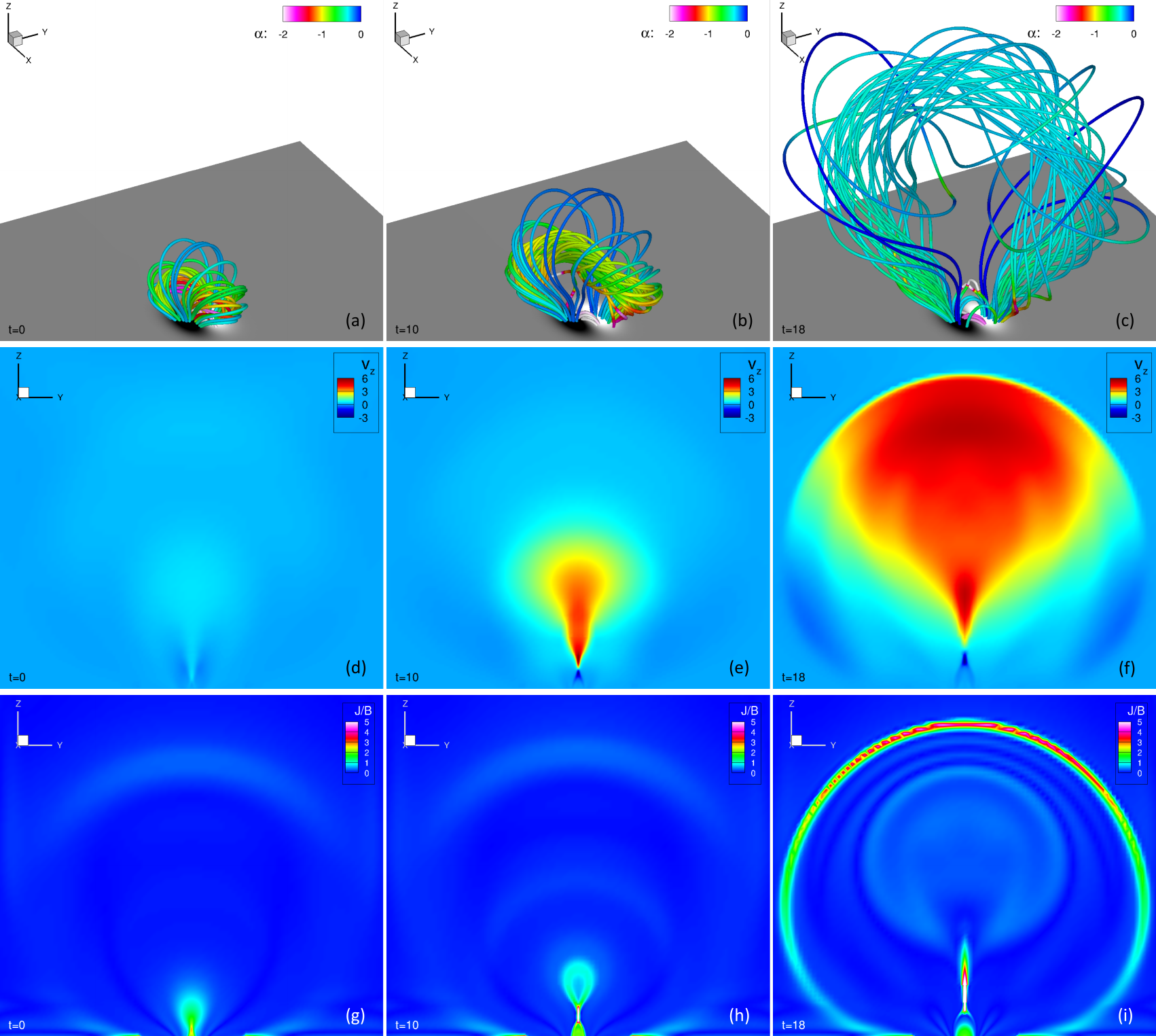}
    \caption{The eruption process in the simulation. (a)-(c) the
      evolution of magnetic field lines colored by the force-free
      parameter $\alpha=\mathbf{J}\cdot\mathbf{B}/B^2$. (d)-(f)
      vertical velocity evolution in the $x=0$ plane. (g)-(h) current
      density evolution in the $x=0$ plane.\label{fig:eruption}}
  \end{centering}
\end{figure*}

\begin{figure*}
  \begin{centering}
    \hspace{0cm}\includegraphics[scale=0.8]{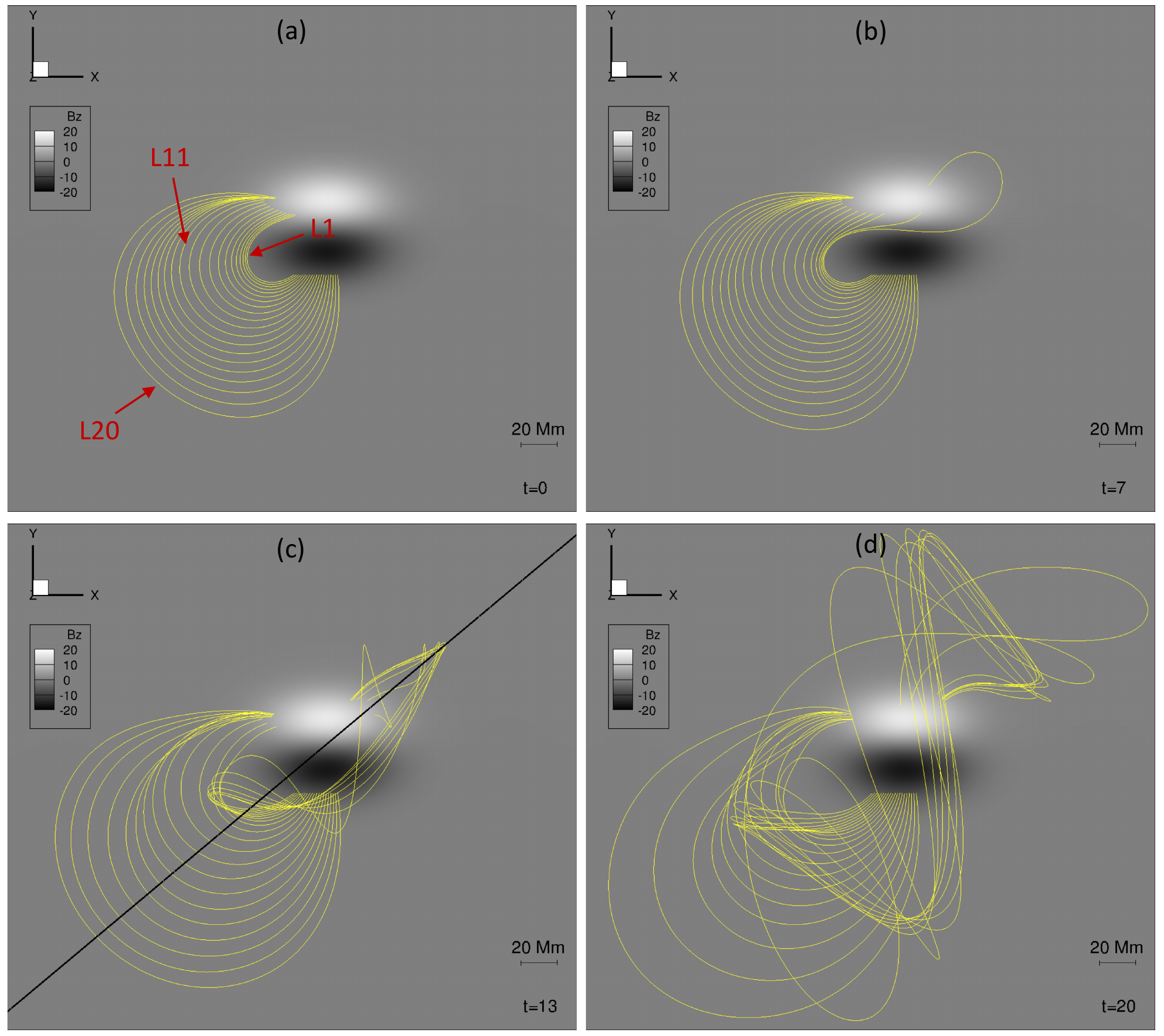}
    \caption{The positions and motion of selected peripheral magnetic
      loops at a face-on state from a top view perspective in the MHD
      simulation. The animated version of this figure is available,
      which runs from $t=0 $ to $t=20$.\label{fig:faceon}}
  \end{centering}
\end{figure*}

\section{Method}\label{sec:method}
The MHD model is aimed to simulate the energy accumulation and
initiation of solar eruption in a simple bipolar active region through
a self-consistent way, which is similar to some previous
studies~\citep[e.g.,][]{ama2003,aul2010}. We solve the isothermal MHD
equations as follows with non-dimensionalization on an adaptively
refining mesh in 3D Cartesian geometry using an advanced conservation
element and solution element (CESE) method \citep{jiang2010}.
\begin{gather}\label{equ}
		\frac{\partial \rho}{\partial t}+\nabla\cdot(\rho \boldsymbol{\rm
		v})=0 \\
		\rho\frac{\mathit{D} \boldsymbol{\rm v}}{\mathit{D} t}=-\nabla
		p+\boldsymbol{\rm J \times B}+\rho \boldsymbol{\rm
			g}+\nabla\cdot(\nu\rho\nabla\boldsymbol{\rm v}) \\
		\frac{\partial \boldsymbol{\rm B}}{\partial
		t}=\nabla\times(\boldsymbol{\rm v}\times\boldsymbol{\rm B}),
\end{gather}
where $\rho$ is the mass density, $\boldsymbol{\rm v}$ the plasma
velocity, $p$
the plasma thermal pressure, $\boldsymbol{\rm J}$ the electric current density,
$\boldsymbol{\rm g}$ the gravitational acceleration, and $\boldsymbol{\rm B}$ the
magnetic field. A small kinetic viscosity $\nu$ in
the momentum Equation (2) is used for the
purpose of keeping numerical stability during the very dynamic phase
of the simulation. The coefficient $\nu=0.05\Delta x^2/\Delta t$
corresponds to a grid Reynolds number of 10 for the length of a grid cell
$\Delta x$. Though the resistivity is set as zero within the
simulation volume, magnetic reconnection can occur via numerical
diffusion when the thickness of a current layer approaches the grid
size. The material derivative $\mathit{D} \boldsymbol{\rm v}/\mathit{D}
t$ in
the momentum Equation (2) represents the time rate of change in the velocity of
a certain plasma element, as it flows along its streamline. The
computational box is $(-24,-24,0)<(x,y,z)<(24,24,48)$,
sufficiently large to minimize numerical influences by side and top
boundaries. The volume is resolved by a nonuniform block-structured
grid with the highest and lowest resolution
$\Delta{x}=\Delta{y}=\Delta{z}=1/16$ and $1/4$, separately. The
parameters for non-dimensionalization related in this paper are
$L_s=16 ~\rm arcsec=11.52 ~\rm Mm$, $t_s=52.5 ~\rm s$,
$v_s=110 ~\rm km~s^{-1}$, $\rho_s=2.29\times10^{-15} ~\rm g~cm^{-3}$
and $B_s=1.86 ~\rm G$ for the length, time, velocity, density and
magnetic field strength scales, respectively.

The initial conditions of the model are as follows. The magnetic field
is potential with a bottom boundary similar to the one in
\cite{ama2003}, mimicking a bipole solar active region with a
relatively strong field gradient near the polarity inversion line
(PIL). The atmospheric plasma is stratified by gravity with no flow,
and the temperature is uniform. To ensure plasma $\beta<1$ in the
lower corona with a minimum value of $5\times10^{-3}$, the local
gravitational coefficient is defined to decrease with increasing
height.

The complete simulation comprises four phases in sequence. We first
quasi-statically energize the initial potential magnetic field through
two vortex flows centred on each polarity at the bottom boundary,
which mimics sunspot rotation during flux emergence and creates
sheared arcade field. Then it is switched to a relaxation phase
without boundary driving. After that, surface diffusion possessing an
enhanced photospheric resistivity is applied to simulate magnetic flux
cancellation observed in the decay phase of active regions
\citep{aul2010}. Finally, when the system becomes unstable and the
field eruption commences, any driving is ceased on the photophore
which holds the magnetic field line-tied. The simulation continues
until any disturbance from the eruption reaches any of the side and
top boundaries. Further details of the model settings, the full
evolution of the four stages, as well as the triggering mechanism and
magnetic topology evolution during the eruption are to be described
elsewhere (Jiang et al. 2021, in preparation).  In this study, we only
focus on the last phase when the magnetic field erupts outwards and,
in particular, the motions of the peripheral magnetic loops associated
with the eruption.

\begin{figure*}
  \gridline{\fig{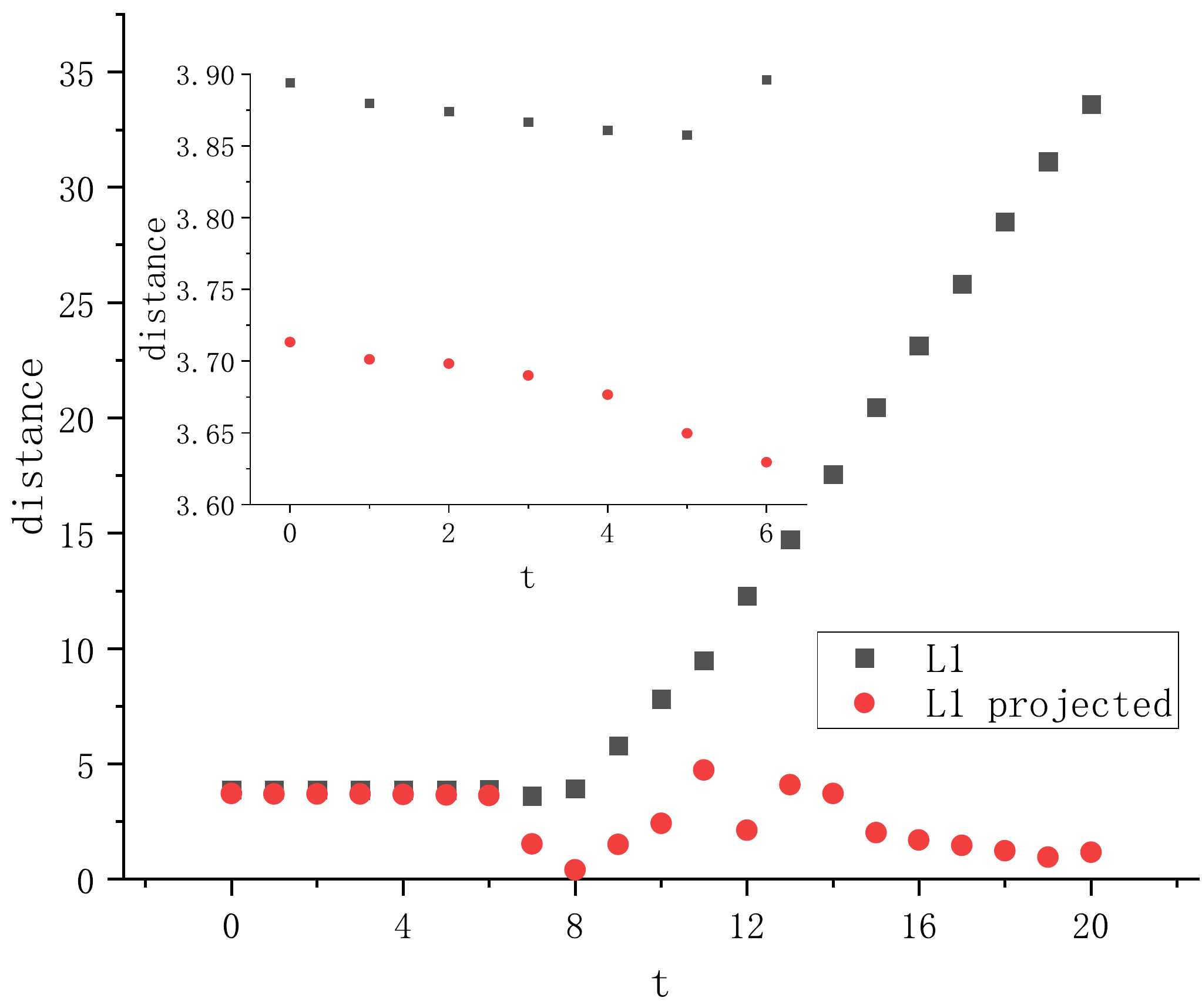}{0.45\textwidth}{(a)}
    \fig{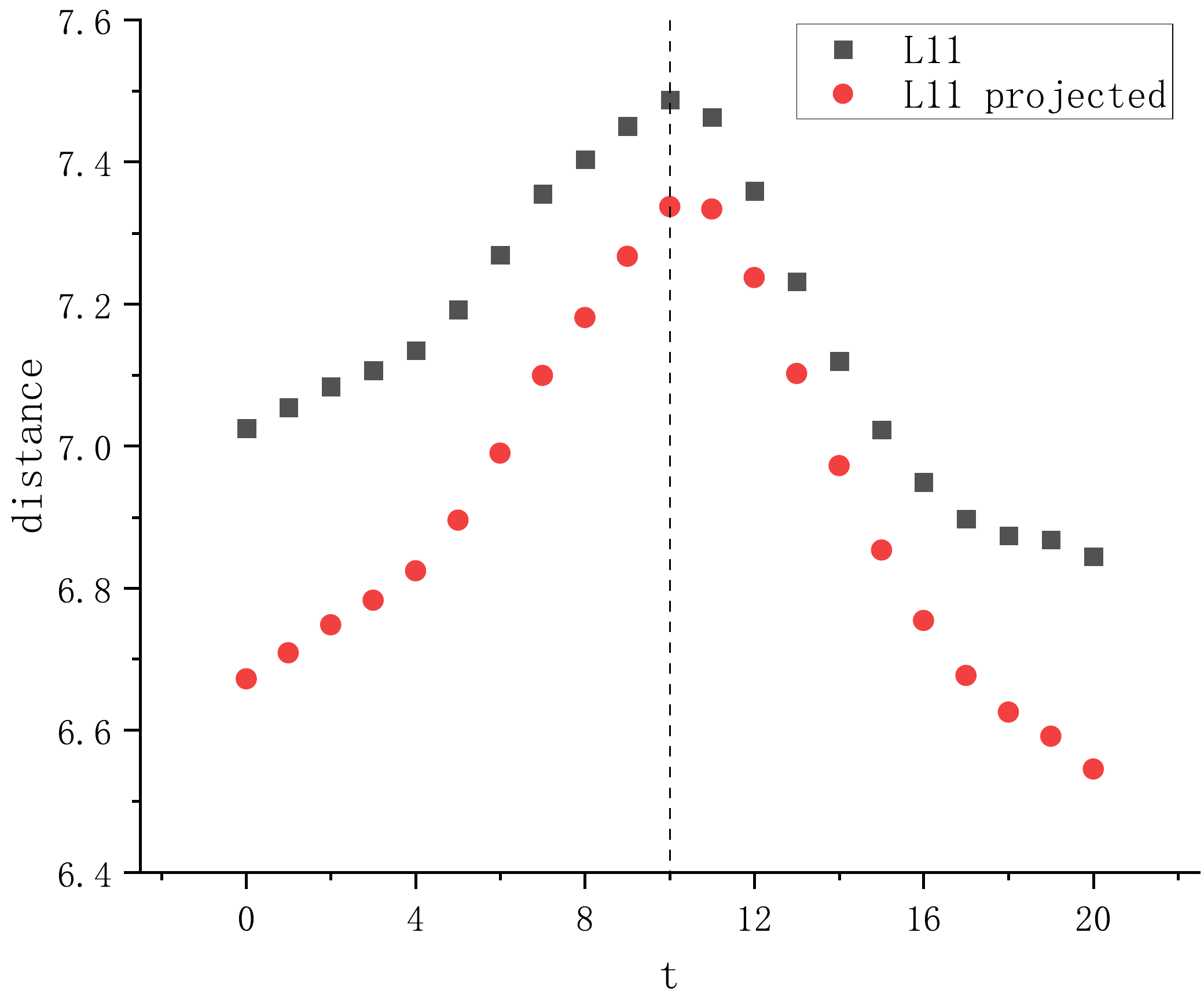}{0.45\textwidth}{(b)} }
  \gridline{\fig{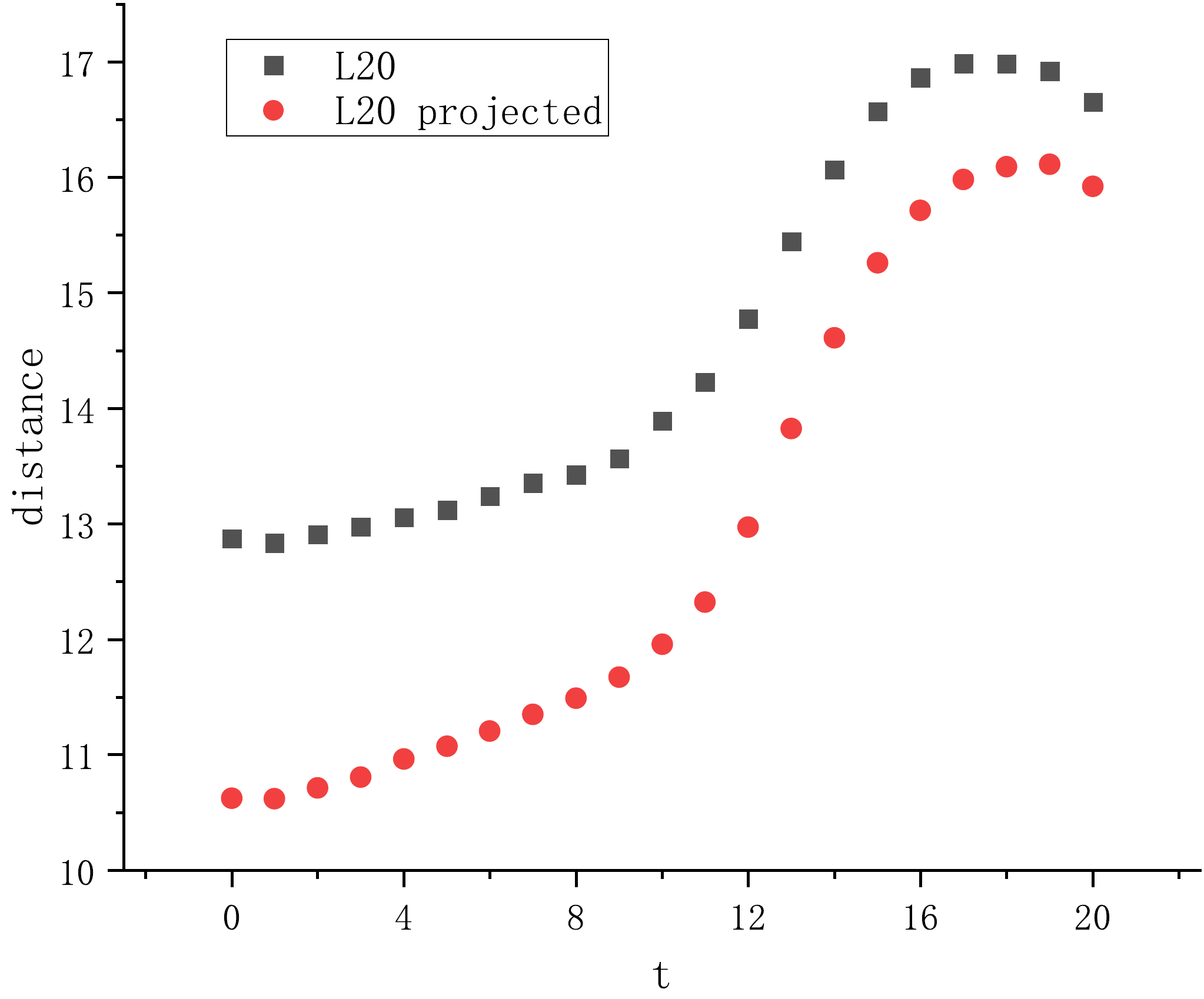}{0.45\textwidth}{(c)}
    \fig{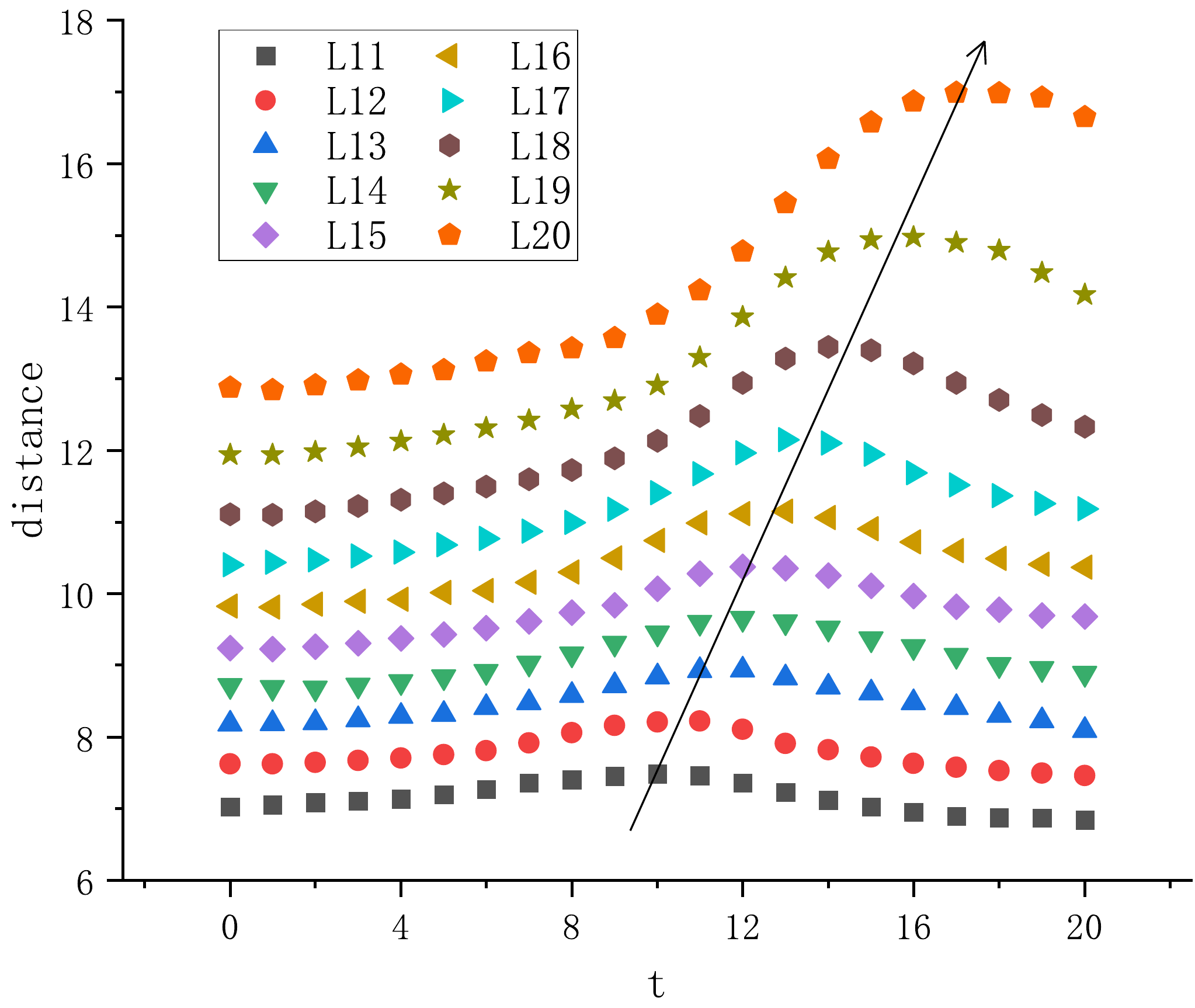}{0.45\textwidth}{(d)} }
  \caption{The detailed kinetics of selected peripheral magnetic
    loops. (a) for the innermost loop L1. The plot is enlarged for the
    data from $t=0$ to $t=6$ in the inset. (b) for the middle loop
    L11. The dashed line shows the transition from expansion to
    contraction. (c) for the outermost loop L20. (d) for L11 through
    L20. The arrow roughly links the inflection points from expansion
    to contraction for these loops.
  \label{fig:motion}}
\end{figure*}

\begin{figure*}
  \begin{centering}
    \hspace{0cm}\includegraphics[scale=0.8]{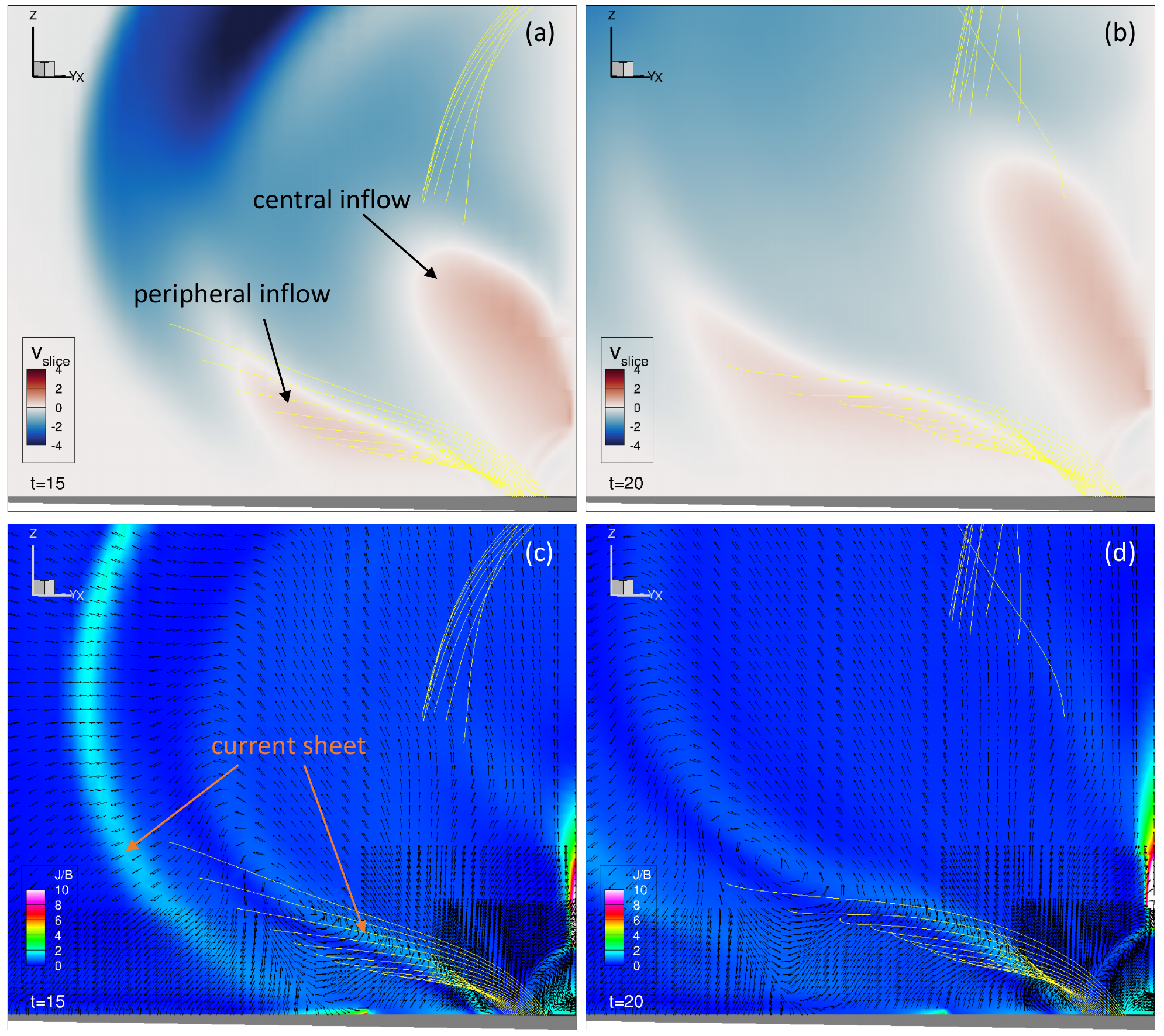}
    \caption{The selected peripheral magnetic loops at an edge-on state. The
    background
      slices are chosen to be the diagonal plane normal to
      $z=0$ whose edge is indicated in
      Figure~\ref{fig:faceon}(c), with (a)-(b) colored by
      horizontal velocity in the slices, and (c)-(d)
      weighted current density. The arrows in (c)-(d)
      shows the velocity vectors in the chosen slices. The
      animated version of this figure is available, which
      runs from $t=0$ to $t=20$.\label{fig:v_j}}
  \end{centering}
\end{figure*}
\begin{figure*}
  \begin{centering}
    \hspace{0cm}\includegraphics[scale=1]{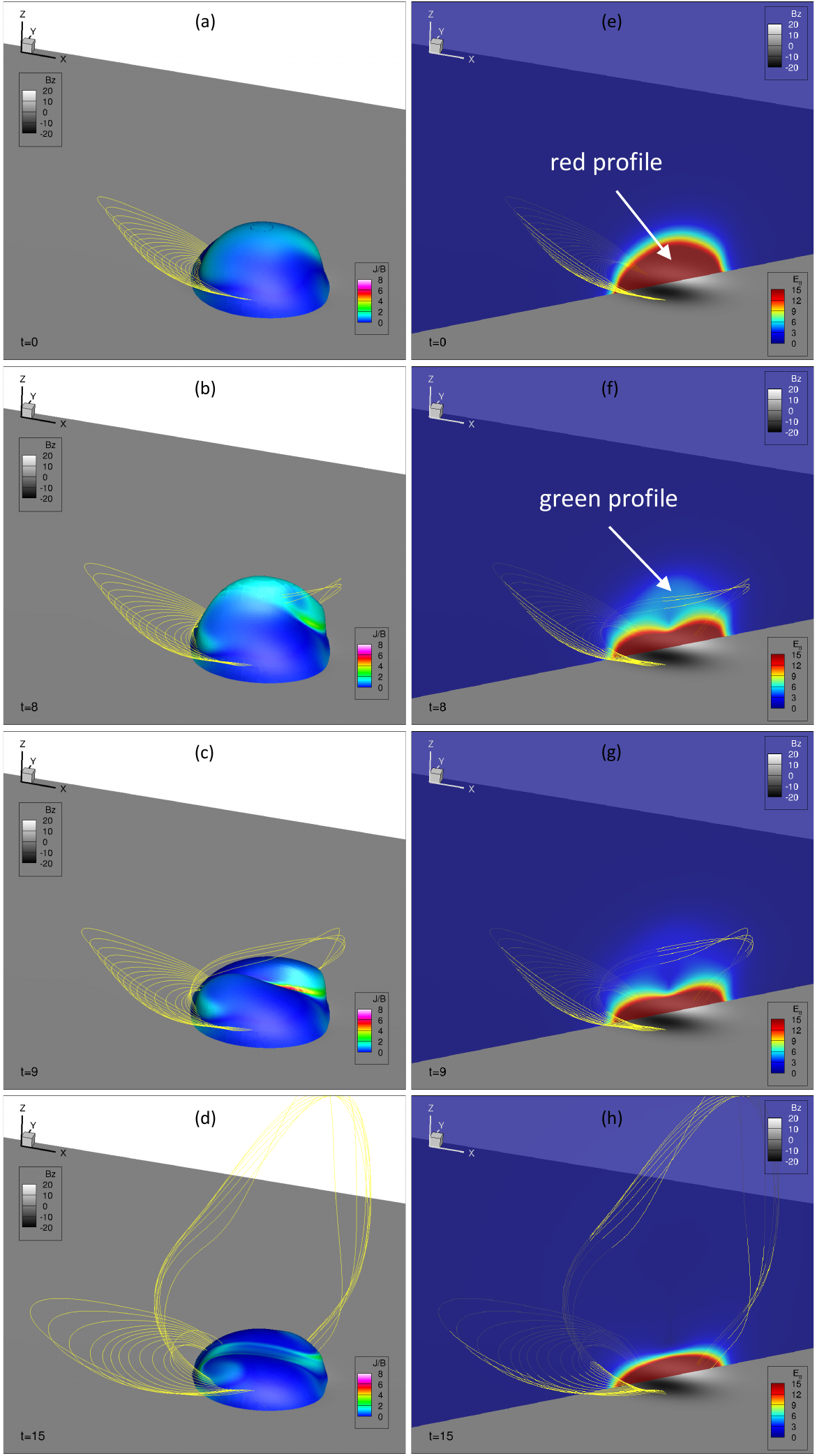}
    \caption{The correlation between peripheral magnetic loop motion
      and central magnetic energy. The isosurface in (a)-(d) denotes
      the magnetic energy density profile $E_B=3.5$, colored with
      weighted current density. (e)-(h) indicates the corresponding
      magnetic energy density profile in the chosen slice as in
      Figure~\ref{fig:faceon}(c). The animated version of this figure
      is available, which runs from $t=0$ to $t=20$.\label{fig:EB2}}
  \end{centering}
\end{figure*}

\section{Results}\label{sec:results}
\subsection{Peripheral Magnetic Loop Motion}\label{sec:motion}

Figure~\ref{fig:eruption} illustrates the evolution of erupting
magnetic field lines, the vertical velocity and current density in the
$x=0$ plane, i.e., the central cross section of the computational
volume in the simulation, which shows a thin current sheet connecting
a erupting plasmoid and a cusp structure of post flare loops below, a
typical process of a coronal mass ejection in observations
\cite[e.g.,][]{sunj2015,wang2016} and many previous numerical
simulations with different scenarios
\cite[e.g.,][]{lin2003,fan2010,tor2018}. In order to study the
peripheral loop dynamics, we select 20 magnetic loops traced from the
negative polarity to study, as shown in Figures~\ref{fig:faceon} with
a face-on view. These loops, possessing minor shear, are anchored at
the periphery of the active region with relatively low magnetic
strength compared with the central area. From $t=0$ to $t=20$, they
behave dynamically when the eruption happens at the center; and
different loops experience different physical processes, which we
discuss in detail in Figure~\ref{fig:motion}. As we can see from the
associated animation attached, all of these peripheral loops
eventually contract apparently towards the central polarities. In
Figures~\ref{fig:v_j} and the associated animations, with an edge-on
perspective, the loop contraction instead of just inclination is
evident. From Figures~\ref{fig:faceon}(a) and (b), we note that the
peripheral contracting loops reside in a second inflow area
($V_{\rm slice}>0$, where $V_{\rm slice}$ represents the velocity in the
chosen slice in Figure~\ref{fig:faceon}(c), and red color indicates
direction towards right, blue towards left), separated from the
traditional inflow region in the center. Figures~\ref{fig:v_j}(c) and
(d) also show that the shrinking loops and the second inflow area are
located between two thin layers of high current density, with the
outer one corresponding to the shock produced by the eruption, and the
inner one related to reconnection between the contracting and erupting
structures, which is discussed in Section~\ref{sec:reconnect}.

In Figure~\ref{fig:motion}(a), (b) and (c), we select 3 representative
magnetic loops from Figure~\ref{fig:faceon}(a), that is, the innermost
loop notated by ``L1", the 11th loop in the middle ``L11", and the
outermost loop ``L20", respectively, to investigate their detailed
kinetics. The black dots in these three figures represent the distance
between the midpoint of the loop and the origin (the center at the
bottom boundary) in the numerical box, while the projected value of
this distance from a top view (i.e., the perspective in
Figures~\ref{fig:faceon}(a)-(d)) is denoted by the red ones. Combined
with Figures~\ref{fig:faceon}(a)-(b) and the associated animation,
Figure~\ref{fig:motion}(a) and its inset show that L1 experiences
running slipping towards the center from $t=0$ to $t=6$, then
reconnects forming a sigmoid, and finally erupts
away. Figure~\ref{fig:motion}(b) presents that before $t=10$, L11
undergoes two expansion phases, with the later one possessing a
relatively higher speed, after which it begins to contract towards the
center, and then at around $t=15$ the contraction starts to
decelerate. Both the real and projected distances at $t=20$ is smaller
than the ones at the beginning time $t=0$, demonstrating a net
contraction for L11, which is consistent with the observation in
\cite{wang2016}. The motion of L20 is shown in
Figure~\ref{fig:motion}(c), which exhibits a similar behavior as L11
but without a full contraction phase due to the termination of the
simulation. We expect that the contraction would proceed if the
simulation continued. It is also noted that the transition from
expansion to contraction for L20 is smoother than that for L11.

Figure~\ref{fig:motion}(d) shows the distances between the midpoints
of loops and the origin for L11 through L20 which do not experience
reconnection during the period of the simulation. The arrow roughly
links the inflection points from expansion to contraction for these
magnetic loops, whose slope indicates an approximate speed
$290~ \rm km~s^{-1}$ with dimension for the outward propagating
contraction signal, comparable to the fast mode speed around the
midpoints of the loops, thus consistent with the prediction in
\cite{wang2018}.

\begin{figure*}
  \begin{centering}
    \hspace{0cm}\includegraphics[scale=0.8]{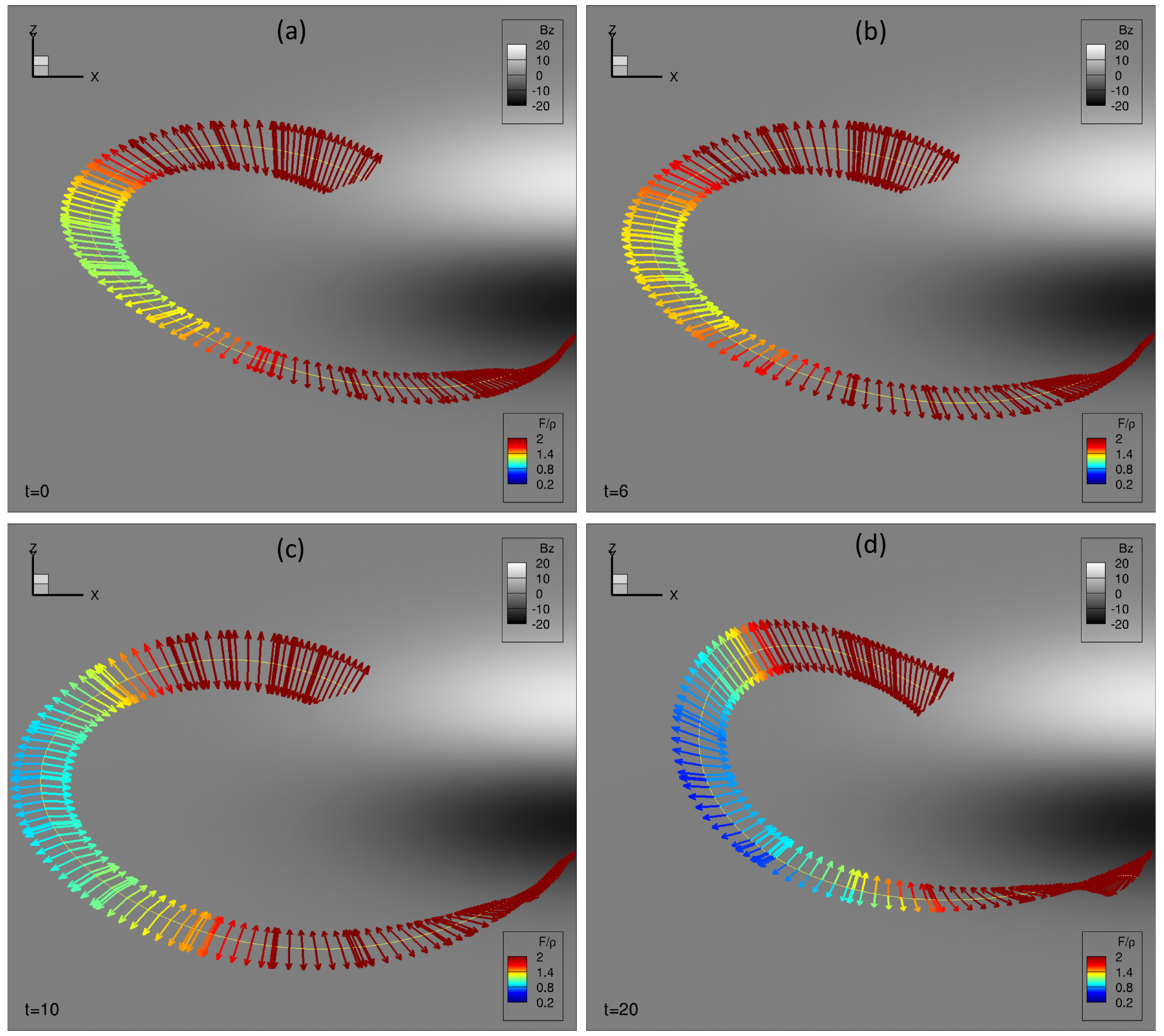}
    \caption{The correlation between peripheral magnetic loop motion
      and Lorentz force components. Outward and inward arrows
      represent MPN and MTN, colored by their magnitudes,
      respectively. The animated version of the figure is available,
      which runs from $t=0$ to $t=20$.\label{fig:L11_dynamic}}
  \end{centering}
\end{figure*}

\begin{figure}
  \begin{centering}
    \hspace{0cm}\includegraphics[scale=1.13]{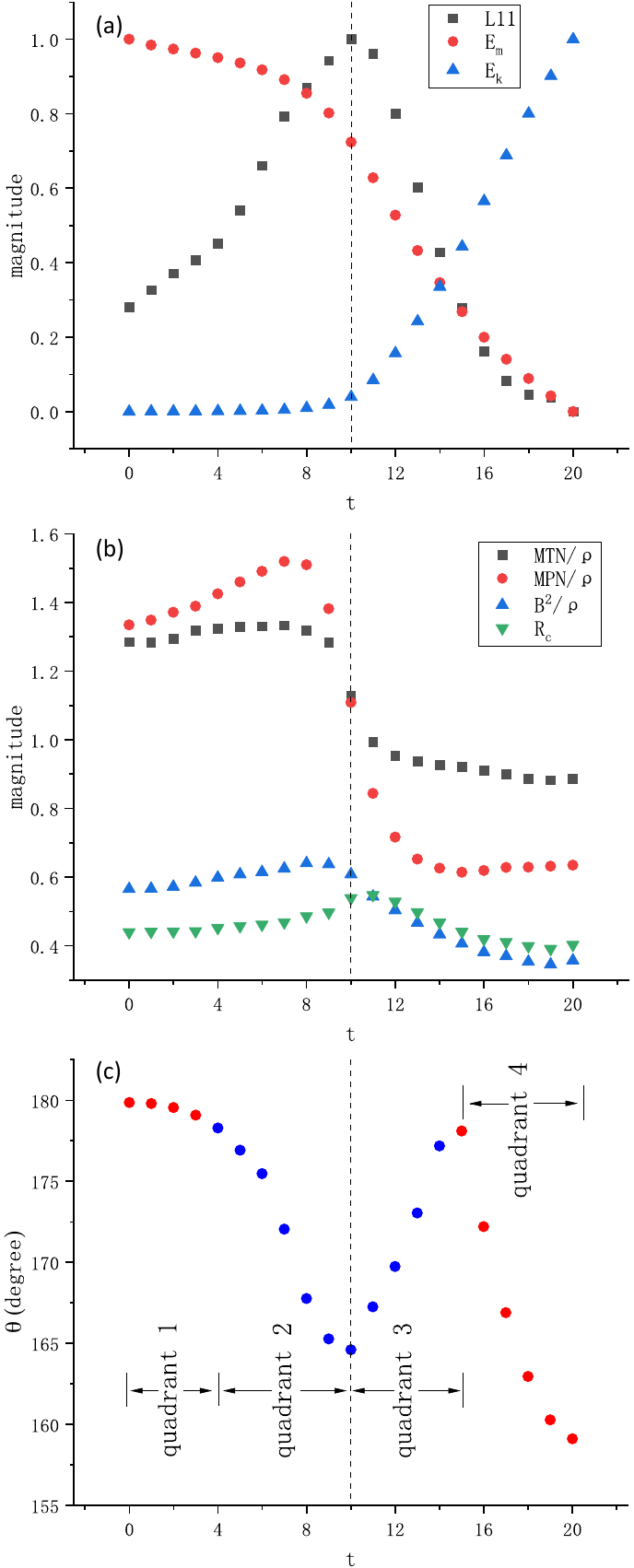}
    \caption{The dynamics of the magnetic loop L11. (a) the evolutions
      of normalized values of L11 distance as in
      Figure~\ref{fig:motion}(b), magnetic energy $E_m$ and plasma
      kinetic energy $E_k$ in the simulation volume. (b) the changes
      in the magnitudes of MTN, MPN, $B^2$ (the three quantities are
      all weighted by the mass density $\rho$), and $R_c$ with
      time. (c) the change in the angle between MTN and MPN with
      time. The dashed line shows the transition from expansion to
      contraction.The dashed line indicates the transition from
      expansion to contraction for L11.\label{fig:L11_force}}
  \end{centering}
\end{figure}

\subsection{Cause of Peripheral Magnetic Loop  Contraction}
\label{sec:reason}

\begin{figure*}
  \begin{centering}
    \hspace{0cm}\includegraphics[scale=0.85]{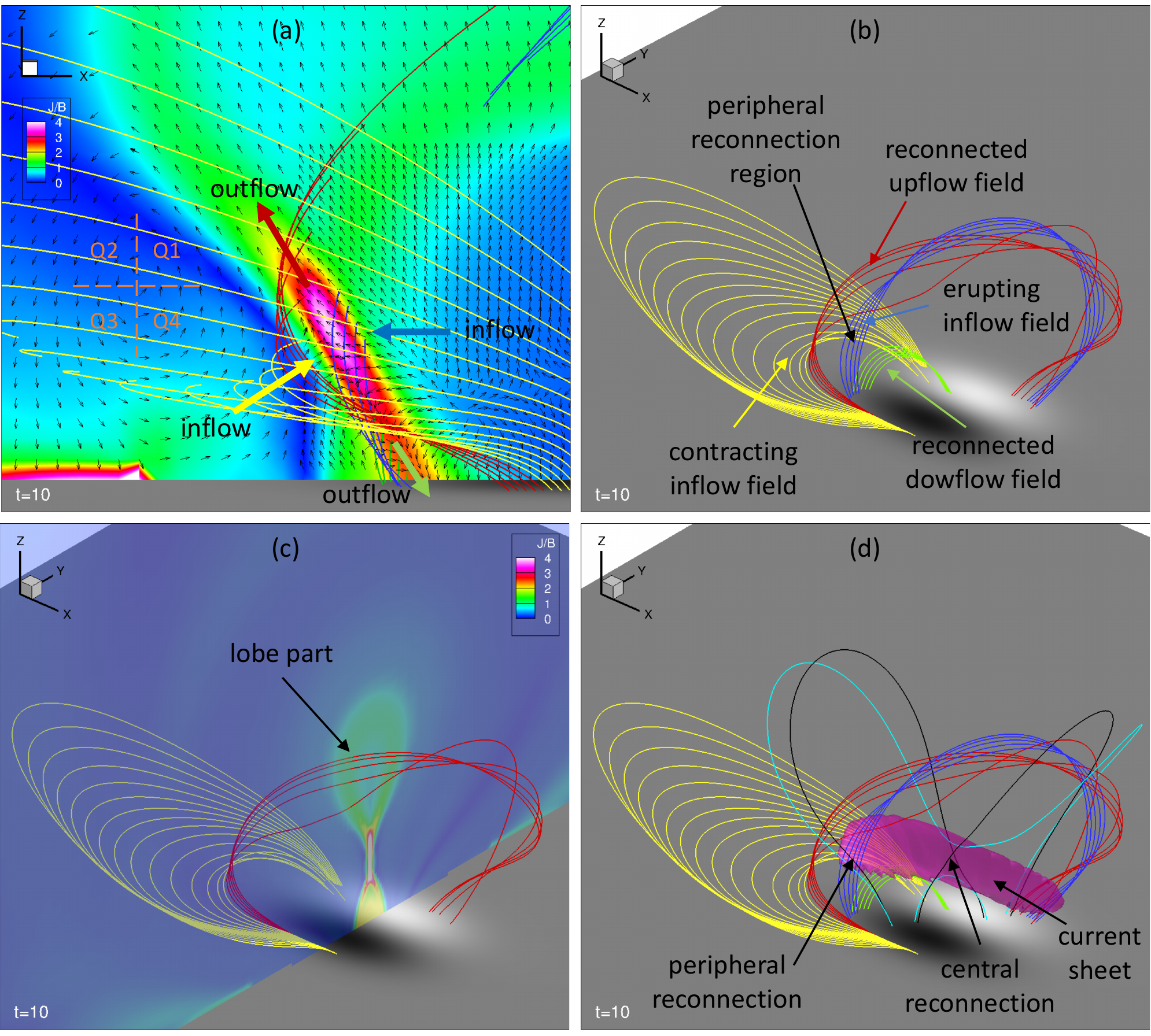}
    \caption{The reconnection between contracting and
      erupting fields in the periphery. (a) vortex flow associated
      with the selected magnetic loops in Figure \ref{fig:faceon}, and
      the peripheral reconnection scenario. The chosen slice is the
      $y=-1$ plane, colored by weighted current density and also added
      with arrows indicating velocity vectors in the plane. $Qn$
      ($n=1,2,3,4$) separated by the two dashed line dividing sections
      of the vortex flow represents quadrant n in Figure
      \ref{fig:L11_force}(c) for simplicity. (b)The peripheral
      reconnection scenario in 3D. (c) shows that the reconnected
      outflowing field lines (red) penetrate through the lobe part of
      the $\Omega$ structure of high current density in the $x=0$
      plane. (d) the relative position of the asymmetric peripheral
      reconnection to the symmetric central tether-cutting
      reconnection. The magenta isosurface denotes the current sheet
      profile of $J/B=3.2$. The animated version of (d) in 3D view is
      available.\label{fig:rec}}
  \end{centering}
\end{figure*}

In Figures~\ref{fig:EB2}(a)-(d) we plot the magnetic energy density
$E_B=B^2/2=3.5$ isosurface colored with normalized current density
$J/B$, to graphically investigate if the contraction of the selected
magnetic loop is qualitatively correlated with magnetic energy release
underneath proposed by \cite{hud2000}. Figures~\ref{fig:EB2}(e)-(h)
show the corresponding magnetic energy density profile in the selected
slice as in Figure~\ref{fig:faceon}(c). From $t=0$
(Figure~\ref{fig:EB2}(a)) to $t=8$ (Figure~\ref{fig:EB2}(b)), the
selected isosurface experiences a slight inflation in the middle
portion (the same as the evolution of the green profile from
Figure~\ref{fig:EB2}(e) to Figure~\ref{fig:EB2}(f)), possibly due to
magnetic energy accumulation as a result of gradual inflows from
either side of the central tether-cutting reconnection (whose position
can be viewed later in Figure~\ref{fig:rec}(d)) in the preflare
phase. During this period, the peripheral loops experience
pre-eruption slow expansion except that the inner ones have
reconnection which may cause the enhanced current in the inflating
middle portion of the isosurface. Though the red profile descends
possibly caused by the gradual central tether-cutting reconnection
dissipation, the global magnetic structure does not erupt, which can
be seen from the attached animation and also from the plasma kinetic
energy later in Figure~\ref{fig:L11_force}(a).  At around $t=9$
(Figure~\ref{fig:EB2}(c)) the isosurface suddenly collapses downward
(corresponding to the collapse of the green profile in
Figure~\ref{fig:EB2}(g)), which could be caused by a runaway
dissipation of magnetic energy in the central reconnection just
beneath the reconnected inner peripheral loops. The accelerated
dissipation of magnetic energy at around $t=9$ can be clearly seen
later in Figure~\ref{fig:L11_force}(a). After that (see
Figure~\ref{fig:EB2}(d) and the associated animation), the reconnected
inner peripheral loops drastically erupt outward; the outer ones
expand significantly, which may be resulted from enhanced magnetic
pressure higher up by the lateral expansion of erupting structures;
and the ones in between sequentially contract into the
magnetic-energy-releasing core. The drastic increase in plasma kinetic
energy after around $t=9$ can also be seen later in
Figure~\ref{fig:L11_force}(a). Thus the expansion and contraction of
the peripheral loops are well correlated with the accumulation and
dissipation of magnetic energy in the central region.

We then focus on the representative dynamics of L11 in
Figures~\ref{fig:L11_dynamic}. Because the
Alfv\'{e}n frozen flux theorem applies in the coronal MHD environment
\citep{pri2014}, the
magnetic field
changes as if it moves with the plasma, thus when we trace the magnetic loop,
the material derivative in the momentum Equation (2) is considered for the
plasma dynamics. On the right hand side of the momentum Equation (2), Lorentz
force dominates in the low-$\beta$ coronal condition. Lorentz force can be
decomposed
into two components, inward magnetic tension force and outward
magnetic pressure gradient force, while the parallel components to the
magnetic field line of these two forces cancel with each other
~\citep{pri2014}, we thus only pay attention to their normal
components to the magnetic field line, abbreviated as MTN and MPN
hereafter. In Figures~\ref{fig:L11_dynamic} we have them plotted for
L11 colored by their magnitudes using the same colorbar ($F/\rho$
represents MTN or MPN weighted by density). At $t=0$
(Figure~\ref{fig:L11_dynamic}(a)), MTN and MPN are well balanced. Then
MPN increases to be larger than MTN as L11 expands
(Figure~\ref{fig:L11_dynamic}(b)). At $t=10$
(Figure~\ref{fig:L11_dynamic}(c)), they both reduce significantly but
MTN starts to be larger than MPN until the end of the simulation
(Figure~\ref{fig:L11_dynamic}(d)), which results in the considerable
contraction of L11.

To be more specific and quantitative, we select the midpoint of L11 to
investigate the evolution of MTN and MPN on it. The evolutions of
normalized values of L11 distance as in Figure~\ref{fig:motion}(b),
magnetic energy $E_m$ and plasma kinetic energy $E_k$ in the
simulation volume are added in Figure~\ref{fig:L11_force}(a) for
reference. Together with MTN and MPN, magnetic energy density
represented by $B^2$ and curvature radius calculated as
$R_c={B^2}/{\rm MTN}$ \citep{pri2014} at the midpoint of L11 are also
illustrated in Figure~\ref{fig:L11_force}(b).  MTN, MPN, and $B^2$ are
all weighted by density $\rho$, which can be considered as the
quantities averaged on each particle. We note that magnetic tension
force and magnetic pressure gradient force are not necessarily
antiparallel with each other, also for their components, MTN and MPN,
which is not considered in \cite{zuc2017}. Thus it is indispensable to
take the angle between MTN and MPN into account for further
discussion, as shown in Figure~\ref{fig:L11_force}(c). The data for
the angle are also colored red or blue, representing the Lorentz force
in $z$ direction (or the $z$ component of the net force of MTN and MPN
$>0$ or $<0$, indicating upward or downward acceleration. The dashed
line in each figure denotes the inflection point from expansion to
contraction for L11. First, combining Figure~\ref{fig:L11_force}(a)
and Figure~\ref{fig:L11_force}(c), we can separate the motion of L11
into 4 quadrants. For example, from $t=0$ to around $t=4$, L11 expands
(Figure~\ref{fig:L11_force}(a)) and also is accelerated upwards
(Figure~\ref{fig:L11_force}(c)), thus categorized as quadrant 1, while
in quadrant 2 L11 is accelerated downward as expanding. Quadrants 3
and 4 are also deduced in the same way. This evolution of motion of
L11 is just a vortex flow proposed by \cite{zuc2017}, which can also
be seen later in Figure~\ref{fig:rec}(a).

Figures~\ref{fig:L11_force}(b) and (c) show that at $t=0$ MTN and MPN
balance well with their magnitudes close and angle antiparallel, as
mentioned earlier for Figure~\ref{fig:EB2}(e). Then significant
enhancement of MPN over MTN until $t=7$ causes L11 \textbf{to be} pushed
outward by the expanding reconnected structure in the
center. Meanwhile, as the central reconnected structure inflates
continuously, the direction of MPN changes from upward and outward to
downward and outward, which makes the decrease in the angle between
MTN and MPN, and the transition from upward to downward of the z
component of the total Lorentz force. This is consistent with the
motion of L11 from quadrant 1 to quadrant 2 in the vortex flow. During
this period, MTN changes little because $B^2$ increases almost in
proportion to $R_c$. Though the magnetic energy in the whole
simulation volume decreases from $t=0$ to $t=7$
(Figures~\ref{fig:L11_force}(a)), it does not make the outward MPN on
L11 decrease because the green magnetic energy profile inflates while
the magnetic energy dissipation mainly occurs in the very center
associated with the central tether-cutting reconnection, as can be
seen from Figure~\ref{fig:EB2}(e) to Figure~\ref{fig:EB2}(f). At
around $t=8$, both MTN and MPN begin to reduce, and the reductions
become significant at around $t=9$, which we suppose is induced by the
runaway magnetic energy dissipation in the central tether-cutting
reconnection as presented in Figure~\ref{fig:EB2}(c) and
Figures~\ref{fig:L11_force}(a). The drastic eruption then follows as
seen from the kinetic energy evolution in
Figures~\ref{fig:L11_force}(a) (also from Figure~\ref{fig:EB2} and the
associated animations). MTN can be reduced because after around $t=8$
$R_c$ grows faster while $B^2$ begins to decrease. This conforms to
physical intuition since the central magnetic liberation could make it
easier
for peripheral loops to become straight and relaxed. The more
drastic reduction in MPN than in MTN leads to the total Lorentz force
starting to be inward and downward at around $t=10$, thus generating
the contraction motion of L11 in quadrant 3 of the vortex flow, which
is contrary to the suggestion by \cite{zuc2017} that the enhancement
of MTN in excess of MPN in the expansion phase finally causes the loop
to contract. The analysis supports the idea that it is the reduction
of magnetic energy of the core region which reduces the outward
magnetic pressure resulting in the contraction of peripheral magnetic
loops conjectured by \cite{hud2000}. As can be imagined, when the
reconnected core structure erupts away, the downward component of MPN
is also reduced, thus making the angle between MTN and MPN
continuously recovered in quadrant 3.

At around $t=15$, L11 steps into quadrant 4 of the vortex flow. Though
at this moment the magnitudes of MTN and MPN become stable, the
direction of MPN begins to incline upward and inward as checked in the
data, which results in the decrease in the angle between MTN and MPN
again and the total Lorentz force pointing upward and inward. This is
because that there is a strong attraction from a current sheet of low
magnetic pressure created by the reconnection between the peripheral
contracting and the central erupting structures, which will be
discussed in the next section.

\subsection{Reconnection between Contracting and Erupting Structures}
\label{sec:reconnect}

Figure~\ref{fig:rec}(a) shows the reconnection between the peripheral
contracting magnetic loops (yellow field lines) and the central
erupting magnetic structures (blue field lines). They act as inflows
into a current sheet created between them. It has also been pointed
out, \textbf{in our discussion of Figures~\ref{fig:v_j}(a) and (b) in Section~\ref{sec:motion},}
that the shrinking loops can be a second inflow in addition to the
traditional central inflow. Red field lines are the reconnection upflow, and green ones the
reconnection downflow. We can see this scenario more clearly in 3D in
Figure~\ref{fig:rec}(b). Because of the reconnection the identity of
the original contracting loops would disappear. It thus could be one
possible explanation for the disappearance of the shrinking loops in
the late phase of contraction from all AIA wavebands in observations.

It is shown in Figure~\ref{fig:rec}(c) that the reconnected peripheral
loops (red field lines) pass through part of the lobe of the $\Omega$
structure of high current density, different from the standard
``CSHKP" flare model where the upflowing structure reconnected from
the central current sheet constitutes the lobe part. We also note in
Figure~\ref{fig:EB2}(d) that the reconnected peripheral loops are
anchored at the hook part of the J-shaped ribbon, consistent with the
discovery of \cite{aul2019}. Figure~\ref{fig:rec}(d) and the animation
attached show the relative position of the peripheral reconnection to
the central one. It can be seen that the peripheral reconnection is a
product of the transition of the reconnection structure along the
current sheet (magenta isosurface) from symmetric in the center
towards asymmetric at the periphery; and it is just located at the
elbow extension part of the inverse S shaped current sheet.

\section{Discussion and Conclusions}
\label{sec:conclusions}

We have exploited a 3D, fully MHD simulation to study peripheral loop
contraction and disappearance in solar events. The inner ones of the
selected magnetic loops at the periphery are found to slip into the
central magnetic energy liberation region and to reconnect forming a
sigmoid structure. This behavior has not been observed in previous
implosion studies since these inner structures are sometimes invisible
by the instruments used \citep[e.g.,][]{wang2016} or too close to the
flaring core \citep[e.g.,][]{sim2013}. The middle and outer loops
selected expand and then contract in sequence from inner to outer
towards the central erupting region, substantiating the reality of the
contraction motion rather than just being a projection effect of
inclination. The transition from expansion to contraction for some
peripheral loops in the simulation is consistent with several events
\citep{wang2016,wang2018}, though it is also found in observation that
large scale loops can directly contract without the expansion phase at
first
\citep{gos2012,liu2012,sun2012,sim2013,yan2013,kus2015,pet2016,wang2018}.

From either graphical in Figure~\ref{fig:rec}(a) or quantitative
inspection in Figure~\ref{fig:L11_force}(c), the contraction following
expansion of the peripheral loops constitutes a vortex flow of four
quadrants, which is consistent with the discovery in \cite{zuc2017}
and \cite{dud2017}. However, the physical cause of the loop shrinkage
is found to be different from their proposal introduced in Section
\ref{sec:intro}. First, the loop motion at the periphery correlates
well with magnetic energy release in the center, and controlled by its
induced changes in two Lorentz force components, as illustrated in
Figure~\ref{fig:EB2}. Then through detailed analysis in
Figure~\ref{fig:L11_force} of the changes in physical quantities, MTN,
MPN, $B^2$, $R_c$, and the angle between MTN and MPN, we conclude that
it is the drastic reduction in MPN by the accelerated central magnetic
energy dissipation that causes the peripheral loop to reverse motion
to contract, in support of the conjecture by \cite{hud2000}. In
contrast to the prediction in \cite{zuc2017}, the magnetic tension
force MTN enhances little during the loop expansion, because $B^2$
varies almost in proportion to the curvature radius $R_c$. MTN is also
found to be reduced as the central magnetic energy dissipation rapidly
intensifies because with less press from the center, magnetic loops
tend to become straight and relaxed, conforming to physical intuition,
but the much more decrease in MPN than in MTN finally leads to the
contraction of magnetic loops in the periphery.

When the peripheral loops continue to contract and enter into quadrant
4 of the vortex flow, they move inward and upward to reconnect with
the central erupting structures, which is the reason they behave as a
second inflow embedded between two current sheets. The reconnection
would destroy the original identities of the shrinking loops, which
thus naturally provides a possible explanation for their disappearance
observed at the periphery in the late phase
\citep{liu2009,sun2012,sim2013,yan2013,kus2015,pet2016,wang2016,wang2018}. It
should be noted that other possibilities, like plasma draining,
cooling or heating, could still remain in different situations, which
needs further investigations including both simulations and
observations.

The peripheral reconnection has also been discovered by \cite{aul2019}
in simulation recently, and verified by following observations
\citep{zem2019,lor2019,dud2019}. It is referred to as ``ar-rf
reconnection'' by the authors, where ``a'' denotes arcade, ``r'' flux
rope, and ``f'' flare loop. This kind of reconnection occurring at the
periphery is an attribute of 3D reconnection geometry, which can
result in drifting of erupting flux rope footpoints. Our work shows
that the involved arcade in the ``ar-rf reconnection'' could be the
contracting loops in the returning part of the vortex flow created at
the periphery.

In this simulation, the peripheral reconnection, located at the elbow
of the inverse S shaped current sheet, is the asymmetric extension of
the symmetric tether-cutting reconnection at the center. The inflow
field from the erupting structure (blue field lines in
Figure~\ref{fig:rec}(d)) for the peripheral reconnection is situated
above both the central current sheet and the concave-upward part of
the long loop (long cyan field line) emanating from the central
reconnection, thus it behaves as constraint for underlying central
eruption. As we can imagine, continuous peripheral reconnection
(between yellow and blue field lines) can make this constraining field
to expand as outflows due to outward enhanced tension of newly
reconnected field lines, which then facilitates the eruption of the
central structure underneath. The central eruption can promote the
vortex flow for the contracting inflow field to be reconnected in the
periphery. Accordingly, the peripheral reconnection and the eruption
of the reconnected core structure form a positive feedback. In term of
results, this is similar to the often quoted positive feedback created
between eruption and central reconnection.

We have to note that the results here are limited to the simulation
and the specified model settings. It is encouraged to use various
numerical methods and configurations to test the conclusions. Further
observations about peripheral implosions will hopefully provide more
actual information.

\acknowledgments J.~W. acknowledges support from the International
Postdoctoral Exchange Fellowship Program, Guangdong Basic and Applied
Basic Research Foundation (2019A1515111008), China Postdoctoral
Science Foundation (2020M670897), National Natural Science Foundation
of China (NSFC 12003005). C.~J. acknowledges support from National
Natural Science Foundation of China (NSFC 41822404, 41731067), the
Fundamental Research Funds for the Central Universities (Grant
No.HIT.BRETIV.201901), and Shenzhen Technology Project
(JCYJ20190806142609035).

\vspace{5mm}


\end{document}